\documentclass[iop]{emulateapj}
\usepackage{graphicx}
\usepackage{amsmath}
\usepackage{amssymb}
\usepackage{natbib}
\usepackage{cases}
\usepackage{paralist}
\usepackage{tablefootnote}
\usepackage{bm}
\usepackage{upgreek}
\usepackage{bbold}
\usepackage{url}
\newcommand{\taurex}{$\mathcal{T}$-REx}

\newcommand{\marple}{{\it Marple}}
\newcommand{\robert}{{\it RobERt}}

\DeclareMathAlphabet{\mathpzc}{OT1}{pzc}{m}{it}

\shorttitle{Dreaming of atmospheres}
\shortauthors{I. P. Waldmann}

\begin{document}

\title{Dreaming of atmospheres}

\author{I. P. Waldmann}
\affil{Department of Physics \& Astronomy, University College London, Gower Street, WC1E 6BT, UK}
\email{ingo@star.ucl.ac.uk}

\begin{abstract}
Here we introduce the \robert~(Robotic Exoplanet Recognition) algorithm for the classification of exoplanetary emission spectra. 
Spectral retrievals of exoplanetary atmospheres frequently requires the preselection of molecular/atomic opacities to be defined by the user. In the era of open-source, automated and self-sufficient retrieval algorithms, manual input should be avoided. User dependent input could, in worst case scenarios, lead to incomplete models and biases in the retrieval. 
The \robert~algorithm is based on deep belief neural (DBN) networks trained to accurately recognise molecular signatures for a wide range of planets, atmospheric thermal profiles and compositions. 
Reconstructions of the learned features, also referred to as `dreams' of the network, indicate good convergence and an accurate representation of molecular features in the DBN. 
Using these deep neural networks, we work towards retrieval algorithms that themselves understand the nature of the observed spectra, are able to learn from current and past data and make sensible qualitative preselections of atmospheric opacities to be used for the quantitative stage of the retrieval process. 
\end{abstract}


\section{Introduction}
The atmospheric retrieval of exoplanetary emission/transmission spectra is a complex undertaking \citep[e.g.][]{2009ApJ...707...24M,Lee:2011gl,2012ApJ...749...93L,Benneke:2013vx,2014RSPTA.37230086G,waldmann15,waldmann15b}. Here, retrieval parameter dimensionality becomes an important factor to consider and though desirable, most times allowing for all known atmospheric species to be fitted is too computationally expensive. Hence, a user defined pre-selection of atmospheric absorbers/emitters must be made. A `seasoned user' would make this pre-selection based on previous experiences and a qualitative recognition of absorption/emission features present in the observed spectrum. Here, the human brain is very good in abstracting previously seen patterns to unseen circumstances, a desirable feature to be replicated by machines . 

As we move to an era of largely automated retrievals, through the provision of open-source code to the community and future ground and space-based spectroscopic surveys, it is important to strive towards universally applicable self-sufficient retrieval algorithms.
In an ideal case scenario, the retrieval suite would posses recognition and learning capabilities similar to the `seasoned user' and would not require any auxiliary user input but the observed spectrum itself. In other words, the program would understand what it is looking at, make a qualitative pre-selection of absorbing/emitting atmospheric species, followed by a quantitative retrieval.  

In \citet{waldmann15}, we began working towards this end by introducing a pattern recognition algorithm, \marple.
Based on principal-component analysis (PCA) facial-recognition approaches, \marple~is able to rapidly sift through large molecular data bases and return a list of the most probable absorbing species in the observed spectrum. This information can then be fed to the \taurex~atmospheric retrieval code \citep{waldmann15,waldmann15b} for a more quantitative analysis. Based on intrinsically linear coordinate transformations, \marple~works well for transmission spectroscopy where the temperature-pressure profile (TP-profile) can be assumed to be isothermal and the transmission approximated by a linear system. 

The emission spectroscopy case is more complicated. Here, the shape of spectral features strongly depends on the varying atmospheric thermal profile as well as varying molecular abundances. Such a non-linear system is often poorly captured by a principal component approach. 

Consequently, we have developed a new neural-network based spectroscopic pattern recognition framework, \robert~(Robotic Exoplanet Recognition), capable of learning and abstracting highly non-linear systems and recognising spectral features found in emission spectroscopy.

In this paper we introduce the concept of deep-belief networks (DBNs) to the recognition of spectral features, describe the training set and algorithm used and discuss \robert's recognition abilities using simulated spectra. 

\section{RobERt}
\label{sec:robert}

\robert~mimics human recognition of spectroscopic features by using a pre-trained deep belief neural network \citep{Hinton:2006bg,Hinton:2007ir,Bengio:2007wn,LeRoux:2010jp,Montavon:2012ii,Bianchini:2014kx} at its core. DBNs are multi-layer non-linear transformations of the input data, in this case the emission spectrum, where each consecutive layer presents a progressively higher level of abstraction of the underlying features in the spectrum.  These levels of abstraction are learned in an unsupervised (i.e. autonomous) fashion from a large catalogue of input spectra. Once these features are learned from the data, a second, supervised learning stage is used to assign the learned features to their correct labels (e.g. H$_{2}$O, CH$_{4}$, etc.). 

Neural networks are now commonly used in complex classification tasks such as image recognition \citep[e.g.][]{2014arXiv1401.5900W, Shen:2015bm,Liu:2014kr,Krizhevsky:2012wl}, speech \& music recognition \citep[e.g.][]{Hung:2005do,Jaitly:2011go,XiaoLeiZhang:2013gy,Pradeep:2014ih}, biology \citep[e.g.][]{1993PhRvE..48.1502H,Plebe:2007hg,Wu:2012wu, Spencer:2015bs} and find increasing use in the classification of galaxies and cosmology \citep[e.g.][]{2004PASP..116..345C,2012MNRAS.424.1409A,2013MNRAS.429.1278K,2014MNRAS.439.2102A, 0004-637X-747-1-59,2015ApJS..221....8H,2016MNRAS.455..370E,2015MNRAS.454.2026D,2015MNRAS.450.1441D}. 

Whereas an in-depth derivation of DBNs is beyond the scope of this paper, we will briefly outline its underlying architecture and implementation. We refer the interested reader to \citet{Bengio:2009kb, Hinton:2012ig} and \citet{Fischer:2014hy} for detailed derivations.

\subsection{Restricted Boltzmann Machines}
\label{sec:rbms}

\begin{figure}
\centering
\includegraphics[width=\columnwidth]{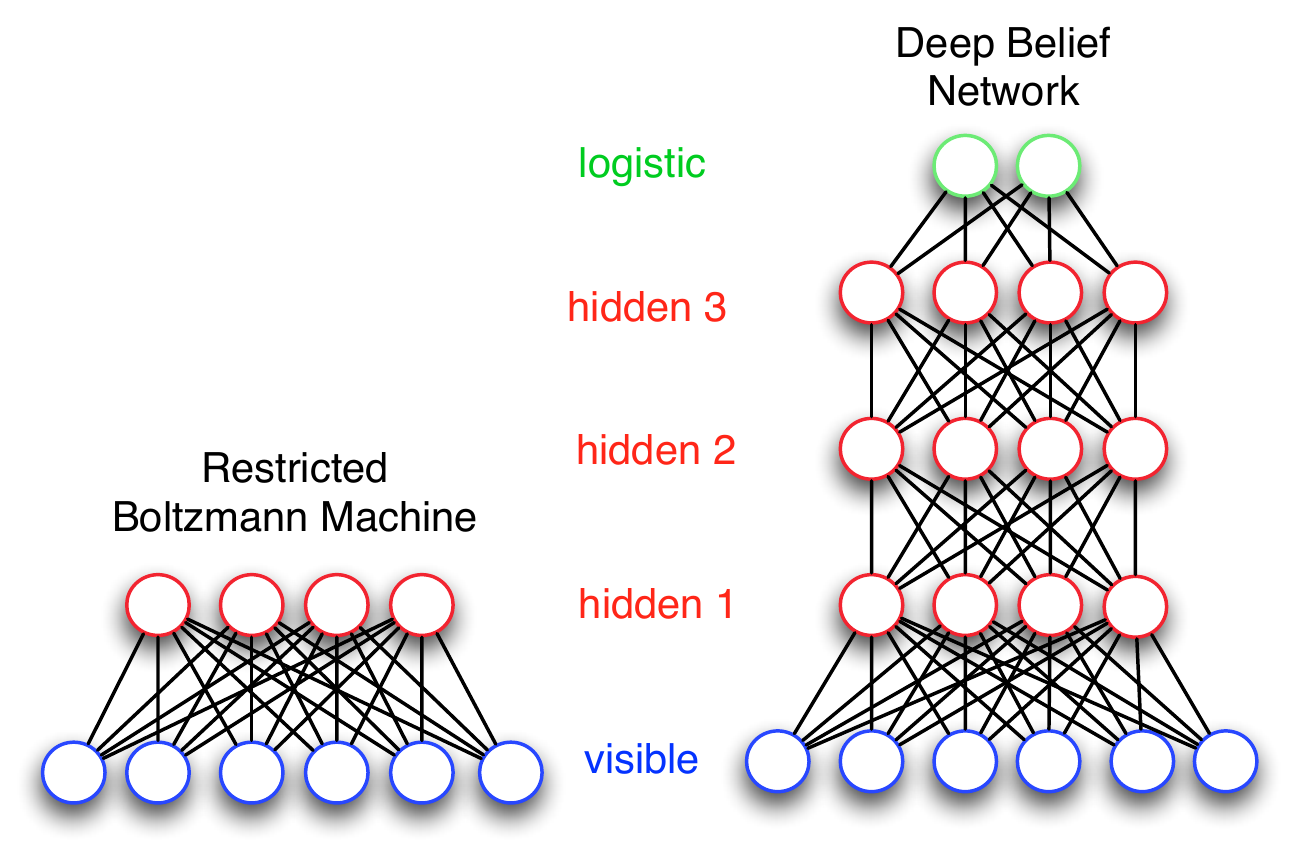}
\caption{Schematic outline of a Restricted Boltzmann Machine (RBM) on the left and a full Deep belief network (DBN) in the form of a Multi-layer Perceptron (MLP) on the right. Blue bottom layer are the `visible units' which are set to the input spectrum during training and recognition. Red are layers of `hidden units' forming increasingly abstract representations of the input layer the further up the network they are. Green represents logistic units linking data labels to the top-layer of hidden units. All units are connected (black lines) with all units in the layers above and below but not intra-level connections exist. It can be seen that the DBN can be built from three consecutive RBMs with the addition of a logistic regression layer.  \label{fig:chart}}
\end{figure}

Figure~\ref{fig:chart} shows a schematic of the the deep belief network. The multi-layer DBN can be constructed from several Restricted Boltzmann Machines \citep{Freund:1992vq,Bishop:2006ui,LeRoux:2008ex, Lee:2011kq, Hinton:2012ig,Bengio:2009kb, 2012arXiv1206.5533B, Montavon:2012ii,Fischer:2014hy} with the addition of a logistic regression layer at the top of the network. The RBM is a two-layer neural network able to learn the underlying probability distribution over its set of input values. It represents a particular kind of Markov Random Field \citep{davison09} consisting of one layer of  binary or Gaussian stochastic visible units (the input data) and one layer of binary stochastic hidden units. In RBMs all hidden units are connected to all visible units but have no intra-layer dependence. Hence all hidden units given the visible units are statistically independent and we can write the probability of all visible units given all hidden units and vice versa as the product of the individual probabilities,

\begin{align}
\label{equ:v-wrt-h}
\begin{split}
P({\bf v} | {\bf h}) &= \prod_{i} P( v_{i} | {\bf h})  \\
P({\bf h} | {\bf v}) &= \prod_{j} P(h_{j} | {\bf v}) 
\end{split}
\end{align}

\noindent where $\bf{v}$ and $\bf{h}$ are the column vectors of visible and hidden units respectively and $i$ and $j$ are their corresponding indices. We now want to find a configuration of the hidden layers, $\bf{h}$, that allows us to reconstruct the input, $\bf{v}$, with minimal error. Since $P({\bf v | h})$ and $P({\bf h | v})$ are factorial, we can write the activation functions of the individual visible and hidden binary units as 

\begin{align}
\begin{split}
P(v_{i} = 1 | {\bf h}) = \varsigma( b_{i} + \sum_{j} h_{j}w_{ij})\\
P(h_{j} = 1 | {\bf v}) = \varsigma( c_{j} + \sum_{i} v_{i}w_{ij})
\label{equ:activation}
\end{split}
\end{align}

\noindent where $\varsigma$ is the sigmoid function

\begin{equation}
\varsigma(x) = ( 1 + e^{-x})^{-1}
\end{equation}

\noindent Assuming both visible and hidden units are binary, RBMs assign an energy term for each configuration of ${\bf v}$ and $\bf{h}$

\begin{equation}
\text{E}({\bf v, h}) = -{\bf b^{T} v - c^{T}h -h^{T}}W{\bf v}
\label{equ:rbm}
\end{equation}

\noindent where ${\bf b}$ and ${\bf c}$ are bias vectors for the visible and hidden units respectively and $W$ is a matrix of connection weights between ${\bf v}$ and ${\bf h}$. The probability over all visible and hidden units $P({\bf v},{\bf h})$ is now given by

\begin{equation}
P({\bf v},{\bf h}) = \frac{e^{-\text{E}({\bf v,h})}}{Z}
\label{equ:free-energy}
\end{equation}

\noindent where $Z$ is the partition function 

\begin{equation}
Z = \sum_{{\bf v,h}} e^{-\text{E}({\bf v,h})}
\label{equ:Z}
\end{equation}

\noindent The probability over the visible units as given by the RBM can now be calculated by summing over all hidden units 

\begin{equation}
P({\bf v}) = \frac{1}{Z}\sum_{{\bf h}}e^{-\text{E}({\bf v,h})}
\label{equ:free-energy}
\end{equation}

\noindent We now train the RBM by finding a set of parameters, ${\bm \theta} = \{ {\bf W},{\bf b}\}$, that maximises the log-likelihood of the data, $ln P({\bf v}| {\bm \theta})$. The derivative of the log-likelihood with respect to the individual weights gives us the gradient on $ln P({\bf v}| {\bm \theta})$

\begin{align}
\frac{\partial ln P({\bf v}| {\bm \theta})}{\partial w_{ij}} &= -\sum_{{\bf h}} P({\bf h} | {\bf v}) \frac{\partial E({\bf v},{\bf h})}{\partial w_{ij}} + \sum_{{\bf v},{\bf h}} P({\bf v},{\bf h})  \frac{\partial E({\bf v},{\bf h})}{\partial w_{ij}} \\
&= \langle v_{i}h_{j}\rangle_{P({\bf h} | {\bf v})} - \langle v_{i}h_{j}\rangle_{P({\bf v},{\bf h})} \\
&= \langle v_{i}h_{j}\rangle_{data} - \langle v_{i}h_{j}\rangle_{model}
\end{align}

\noindent where $\langle v_{i}h_{j}\rangle_{data}$ is the expectation value of all the hidden and visible unit activations given the training data and  $\langle v_{i}h_{j}\rangle_{model}$ is the same expectation under the reconstructed model distribution. The cost function for the optimisation algorithm is now simply given by  

\begin{equation}
\Delta w_{ij} = \epsilon(\langle v_{i}h_{j}\rangle_{data} - \langle v_{i}h_{j}\rangle_{model}) 
\end{equation}

\noindent where $\epsilon$ is a learning rate parameter.  

Training can be performed using simple gradient descent. However, an exact calculation of $\langle v_{i}h_{j}\rangle_{model}$ is highly computationally expensive.
The likelihood gradient can be approximated by sampling the likelihood using Gibbs sampling \citep{Press:2007:NRE:1403886}. Here samples are iteratively drawn from $\langle v_{i}h_{j}\rangle_{data}$ and  $\langle v_{i}h_{j}\rangle_{model}$ until the Markov Chain Monte Carlo (MCMC) sampling converges. Contrastive Divergence  \citep[CD,][]{Hinton:2002ic} further simplifies the Gibbs sampling process by breaking the requirement for exact convergence and restricting the MCMC chain to a few (as few as one) iterations. This leads to significant gains in convergence speed. For an in-depth explanation of CD, we refer the reader to \citep{Hinton:2002ic,NIPS2006_3048, Bengio:2009kb}.

\subsection{Deep Belief Networks}

We now construct the DBN using RBMs as building blocks. As convention and in accordance with figure~\ref{fig:chart}, we refer to the data input to be at the `bottom' of the network and increase in abstraction as we go `up' the network. 

The bottom RBM has the normalised emission spectrum as input (i.e. visible) units. Here a binary representation of the observed data is not ideal and we replace ${\bf v}$ with Gaussian units. These better represent the continuous values found in spectroscopic data. The hidden units and all higher DBN layers remain binary. For the Gaussian RBM layer, the unit activations \citep{Krizhevsky09learningmultiple,WangMW14} become

\begin{align}
\begin{split}
P(v_{i}  | {\bf h}) = \mathcal{N}(v_{i}; b_{i} + \sum_{j} w_{ij}h_{j}, \sigma_{i}^{2}) \\
P(h_{j} = 1 | {\bf v}) = \varsigma( c_{j} + \sum_{i} w_{ij}\frac{v_{i}}{\sigma_{i}^{2}})
\label{equ:activation}
\end{split}
\end{align}

\noindent where $\mathcal{N}$ is the Normal distribution and $\sigma$ the standard deviation of the spectrum. Furthermore, we substitute the energy term (equation.~\ref{equ:rbm}) with

\begin{equation}
\text{E}({\bf v, h}) = \sum_{i} \frac{(v_{i} - b_{i})^{2}}{2 \sigma_{i}} - \sum_{j} b_{j} h_{j} - \sum_{i,j} \frac{v_{i}}{\sigma_{i}} h_{j} w_{i,j}
\label{equ:rbm2}
\end{equation}

\noindent  We now learn the RBM greedily until convergence and take the resulting hidden layer as input to the next higher up RBM. We repeat this process for three consecutive RBMs. This constitutes the unsupervised training stage as the DBN learns on un-labeled data.

Once the RBM layers are trained, we form a Multi-Layer Perceptron (MLP) by attaching a logistic regression layer to the top layer of the network (equation~\ref{equ:activation}). This links the top most hidden units to the data labels (e.g. H$_{2}$O, CH$_{4}$, CO$_{2}$, etc.). We now greedily learn the whole network using stochastic gradient descent by presenting a spectrum of a given composition and its corresponding data label to the network. This supervised learning has two purposes: 1) it fine tunes the network, 2) it associates labels to the network.  
More specifically, in the supervised learning stage, the RBM layers are fixed and act as a feed-forward network. The logistic regression layer now learns the mapping between the high-level representations of the upper RBM layer and the associated data labels. We refer the interested reader to the standard literature \citep[e.g.][]{Bishop:2006ui,Hilbe09} for an in-depth treatment of logistic regression.

We learn the MLP using mini-batch stochastic gradient descent \citep{Li:2014:EMT:2623330.2623612}. Mini-batches determine the number of training examples looked at simultaneously before updating the DBN weights. Looking at `chunks' of data simultaneously, allows us to vectorise the gradient computation and achieve higher convergence speeds than for standard stochastic gradient descent methods.
We did not require the use of any regularisations during supervised learning, but employ `early stopping' criteria to avoid overfitting (see section~\ref{sec:training}). It is worth mentioning that `dropout' algorithms \citep{hinton2012improving,JMLR:v15:srivastava14a} have recently been shown to reach lower reconstruction errors than conventional supervised learning (with or without regularisation) and are found to be highly robust against overfitting, hence avoiding the need for early stopping criteria.


\section{Implementation and training}
\label{sec:setup}

\robert~is written in python using the {\ttfamily scipy} optimisation toolbox and the {\ttfamily theano}\footnote{\url{https://pypi.python.org/pypi/Theano}} library. {\ttfamily Theano} is a very powerful graph and symbolic math toolbox with efficient parallelisation (through the BLAS library) and native GPU support. 
The training data was generated using \taurex~run with OpenMP parallelisation to produce the required grid of emission forward models. 

\subsection{Training data set}

\begin{table}[b]
\centering
\caption{Summary of training set} \label{tbl:training}
\begin{tabular}{| r | l |}
\hline
    No. planets     & 5 \\
    Planets$^{1}$ & WASP-12b, HD189733b, \\
                          &HD209458b, HAT-P-11b, GJ1214b \\
    No. molecules & 10 \\
    Molecules       & H$_{2}$O, HCN, CH$_{4}$, CO$_{2}$, \\
                           &CO, NH$_{3}$, NO, SiO, TiO, VO \\
    Abundance range & $1\times10^{-7} - 1\times 10^{-2}$ \\
    Compositions / planet & 5 \\
    TP-profiles / planet & 7 \\
    $\lambda$ range & 1 - 20$\mu$m\\
    Resolution & 300 (constant) \\
    Points / spectrum & 900\\
    Spectra / planet & 17150 \\
    Spectra total & 85750 \\\hline
 \end{tabular}
\flushleft {\tiny $^{1}$ all parameters from \url{ http://exoplanet.eu}}
\end{table}

In the unsupervised training stage, \robert~requires a large set of example emission spectra to train with. Such a training set should include a broad range of planet types, atmospheric trace gasses and TP-profiles. We considered a total of five planets ranging from warm SuperEarths \citep[GJ1214b,][]{charbonneau09} to the strongly insolated hot-Jupiters \citep[e.g. WASP-12b,][]{Hebb:2009fw}. In total we simulated 17150 emission spectra per planet and 85750 spectra in total. Each spectrum contains only one trace gas species at a time and no mixtures are considered in the training set. Table~\ref{tbl:training} summarises the training set parameters. The creation of the training set took $\sim$3 hours on 96 Intel Xeon E5-2697v2 cpus.  

The data set was now randomly divided into 80$\%$ training data and 20$\%$ test data. \robert~is only trained on the training data with random selection of spectra from the test data presented to \robert~at every $N^{th}$ iteration of the supervised learning to test \robert's prediction accuracy.

\subsubsection{Normalisation}
\label{sec:norm}

\begin{figure}
\centering
\includegraphics[width=\columnwidth]{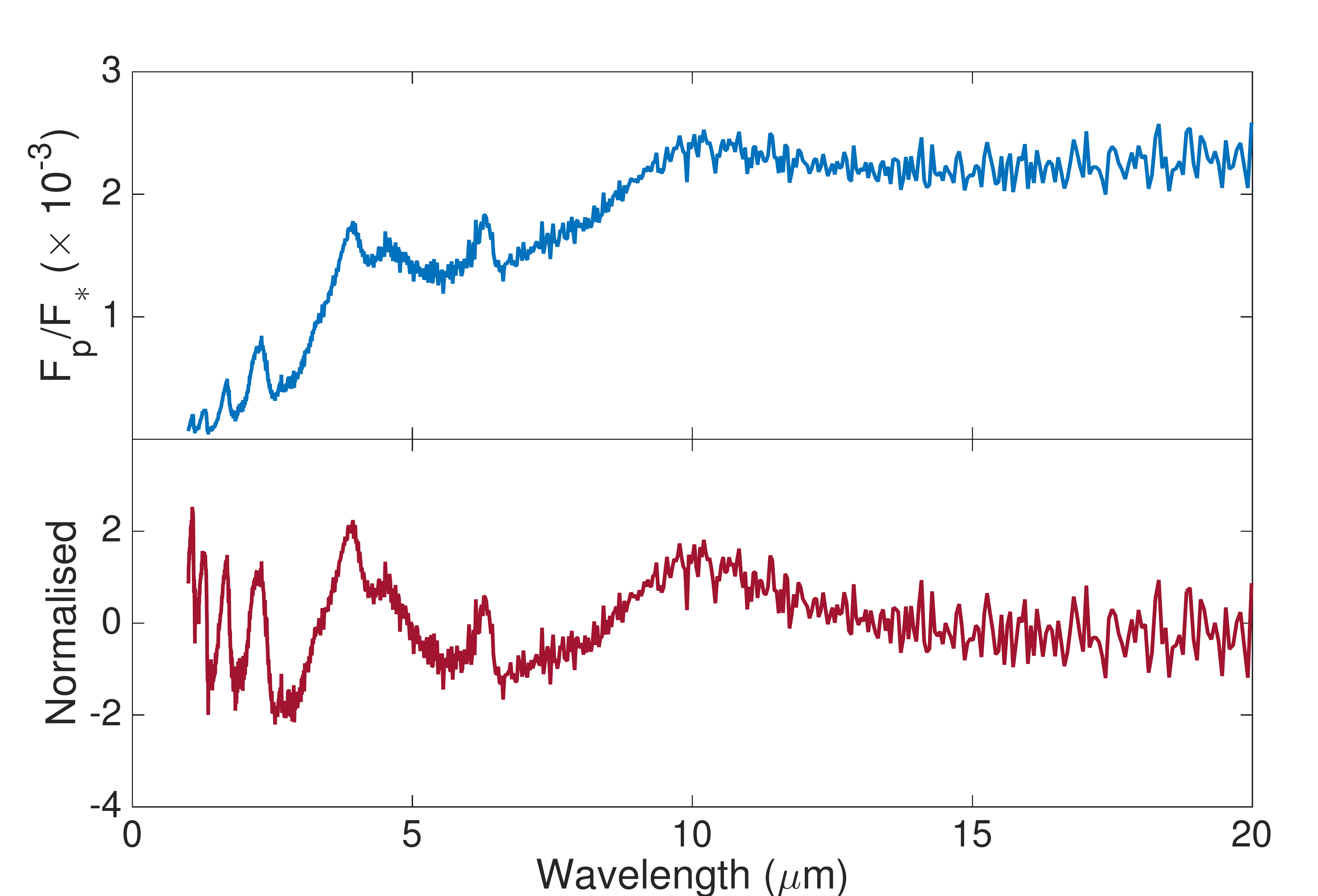}
\caption{Top: example spectrum of a hot-Jupiter (water only) generated by \taurex. Bottom: the normalised emission spectrum used for training \robert. \label{fig:training}}
\end{figure}

Before training \robert~on the catalogue of input spectra, we first normalise the input to a zero mean and unit variance grid. Though this is not strictly necessary, the normalisation significantly improves convergence properties of DBNs. The normalisation consists of three steps:

1) We normalise the emission spectrum with the Planckian of the planet's host star to obtain the planetary intensity 

\begin{equation}
{\bf I}_{p} = ({\bf F}_{p}/{\bf F}_{\ast}) \times {\bf BB}_{\ast} \left ( \frac{R_{\ast}}{R_{p}} \right )^{2}
\label{equ:norm1}
\end{equation}

\noindent where ${\bf F}_{p}/{\bf F}_{\ast}$ is the column vector of the planetary/stellar flux ratio and ${\bf BB}_{\ast}$ is the Planck function at the stellar temperature. This normalisation step ensures that the training process is not biased by the underlying stellar black body function. 

2) We now convert $I_{p}$ into brigthness temperatures using 

\begin{equation}
{\bf T}_{p} = \frac{hc}{{\lambda} k} \times \left [ \text{log}\left ( \frac{2hc^{2}}{{ I}_{p}(\lambda) {\lambda}^{5}} +1 \right ) \right ]^{-1}
\label{equ:norm2}
\end{equation}

\noindent where $k$ is the Boltzmann constant, $h$ the Planck constant, $c$ the speed of light and $\lambda$ the wavelength. 

3) Finally we subtract the mean value of ${\bf T}_{p}$ and normalise to unit variance to give the normalised spectrum~$\hat{\bf T}_{p}$

\begin{equation}
\hat{\bf T}_{p} = \frac{{\bf T}_{p} - \bar{ T}_{p}}{\sqrt{\text{var} ( {\bf T}_{p} - \bar{ T}_{p})}}
\end{equation}

\noindent Figure~\ref{fig:training} shows an example input spectrum of H$_{2}$O before normalisation (top, blue) and after normalisation (bottom, red). 

\subsection{Training}
\label{sec:training}

\robert~is now set up to contain three RBM levels of 500, 200 and 50 neurons from bottom to top respectively, with the input data vector containing 900 spectral points. As discussed in section~\ref{sec:discussion}, we find that slightly smaller networks have similar performance levels but larger networks are too redundant. 

The unsupervised training stage ran over 100 iterations per RBM level at a learning rate of $\epsilon = 0.01$. We find that for all layers, convergence is typically reached between the 80$^{th}$ - 90$^{th}$ iteration. 
During the supervised training stage, we adopt a learning rate of $\epsilon = 0.01$ and a mini-batch sizes of typically 100 training spectra. 
The reconstruction error of the DBN given the test data is computed at each training epoch. Convergence of the supervised learning is reached when no improvement in reconstruction error is obtained over a maximum of 20 epochs and the iteration with the lowest reconstruction error is then taken as final result. This early stopping prevents significant overfitting during the supervised training stage.  

The full training process takes $\sim$1.5h on 6 cpu cores or $<$10 min. using a Nvidia Tesla K40 card (2880 GPUs). 
\robert~completes the supervised training stage with a test data recognition accuracy of 99.7$\%$. 

\section{Recognition of emission spectra}
\label{sec:identification}

\begin{figure}
\centering
\includegraphics[width=1.05\columnwidth]{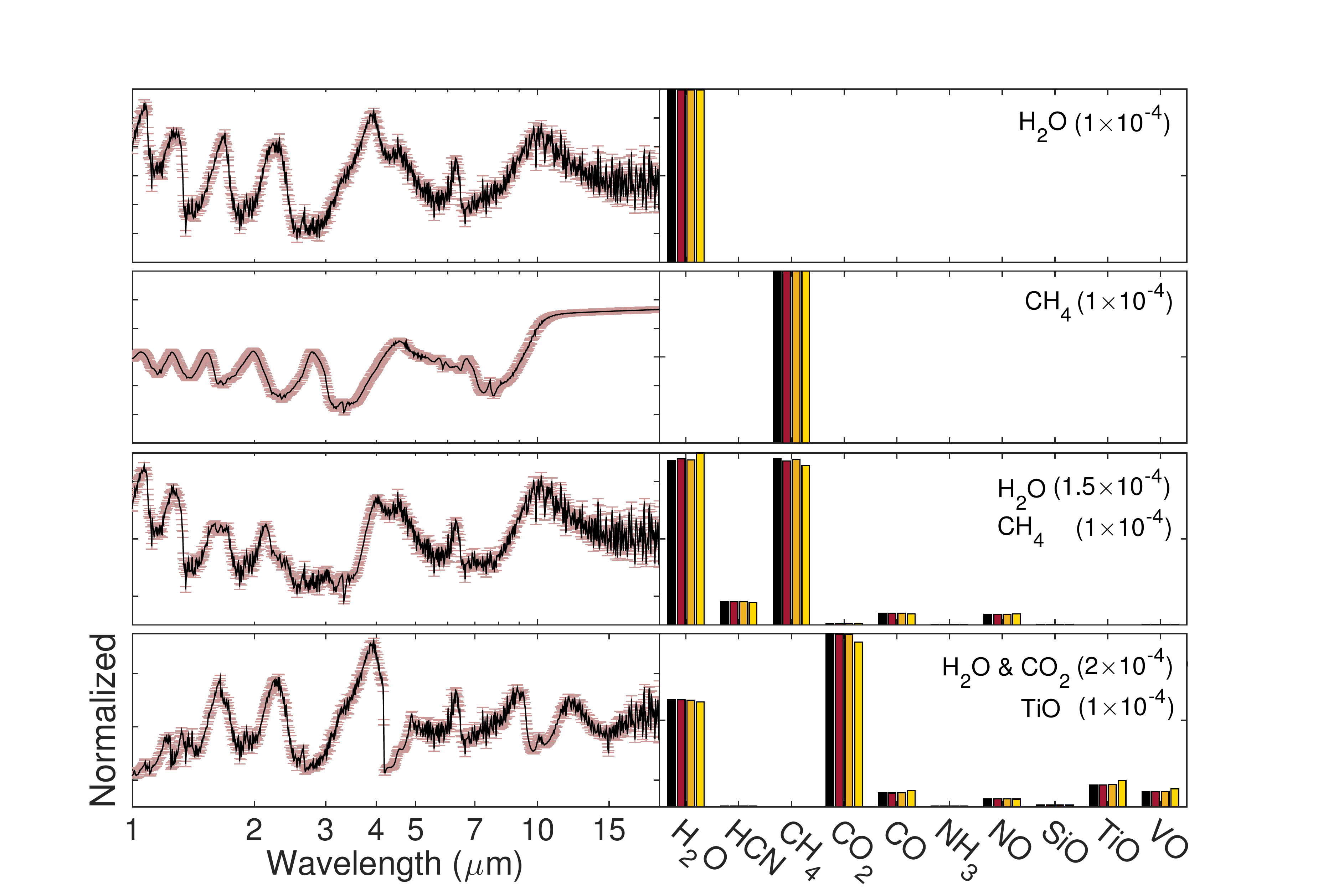}
\caption{Left: example normalised emission spectra at S/N = 20. From top to bottom the spectral compositions are: 1) H$_{2}$O ($1\times 10^{-4}$); 2) CH$_{4}$ ($1\times 10^{-4}$); 3) H$_{2}$O ($1.5\times 10^{-4}$) \& CH$_{4}$ ($1\times 10^{-4}$); 4) H$_{2}$O ($2\times 10^{-4}$), CH$_{4}$ ($2\times 10^{-4}$) \& TiO ($1\times 10^{-4}$). Right: Corresponding probability of the molecule being present in the spectrum to the left. All probabilities are normalised ($p(x) / \text{max}[p(x)]$) for clarity and colour coded to represent 4 different S/N values of the input spectrum: 20 (black), 10 (brown), 5 (orange), 2 (yellow).   \label{fig:identification}}
\end{figure}

One major advantage of DBNs is their ability to generalise patterns over large ranges of parameter spaces, both seen and perviously unseen by the network. To demonstrate this behaviour, we generated emission spectra of the hot-Jupiter WASP-76b \citep{2013arXiv1310.5607W}, unknown to \robert, for a variety of trace gas molecules, mixtures and signal-to-noise (S/N) ratios. 
The spectral recognition process now proceeds in three stages:

1) The observed spectrum is normalised following the steps described in section~\ref{sec:norm}. 

2) The mean of each spectral bin is randomly perturbed within the measurement error bar, resulting in a `noisy' spectrum.  

3) The visible units of the DBN are set to the normalised, noisy spectrum and the DBN is run in the forward direction to obtain the label probabilities $P(label)$. 

Steps~2~\&~3 are repeated 100 times and the label probabilities recorded, summed and normalised. 

Figure~\ref{fig:identification} shows four normalised example spectra and the results of \robert's identification for S/N ratios of 20, 10, 5 and 2. Spectra containing only one main trace gas components are recovered $>$99$\%$ of the time, across all planet types considered. This remains true for strongly saturated spectra with molecular abundances of $> 1\times 10^{-2}$ and very low S/N values. Surprisingly even S/N ratios of 0.5 - 1.0 allow \robert~to recognise the dominant trace gas component with good accuracy. 
\robert~was trained on only individual trace gases, i.e. pure water spectra or pure methane spectra, but not on mixtures of trace gasses. This is mainly due to the very large number of permutations required to represent mixtures of molecules accurately over varying abundances and TP-profiles in the training data. It is hence encouraging to see that \robert~understands mixtures well when presented with them. Figure~\ref{fig:identification} shows two examples of spectra containing H$_{2}$O + CH$_{4}$ and H$_{2}$O, CO$_{2}$ \& TiO. In the three molecules example, \robert~identifies the main constituents, water and carbon-dioxide, with a high probability and the third constituent is either attributed to TiO, VO, CO or NO with TiO having the highest probability of these candidates. In an automated retrieval context, the retrieval code would run a first pass with CO$_{2}$, H$_{2}$O, TiO, VO, CO \& NO as input and proceed to nested model down-selection in subsequent retrieval runs \citep{waldmann15}.

\subsection{Restricted wavelength ranges and resolution miss-matches}

\begin{figure}
\centering
\includegraphics[width=\columnwidth]{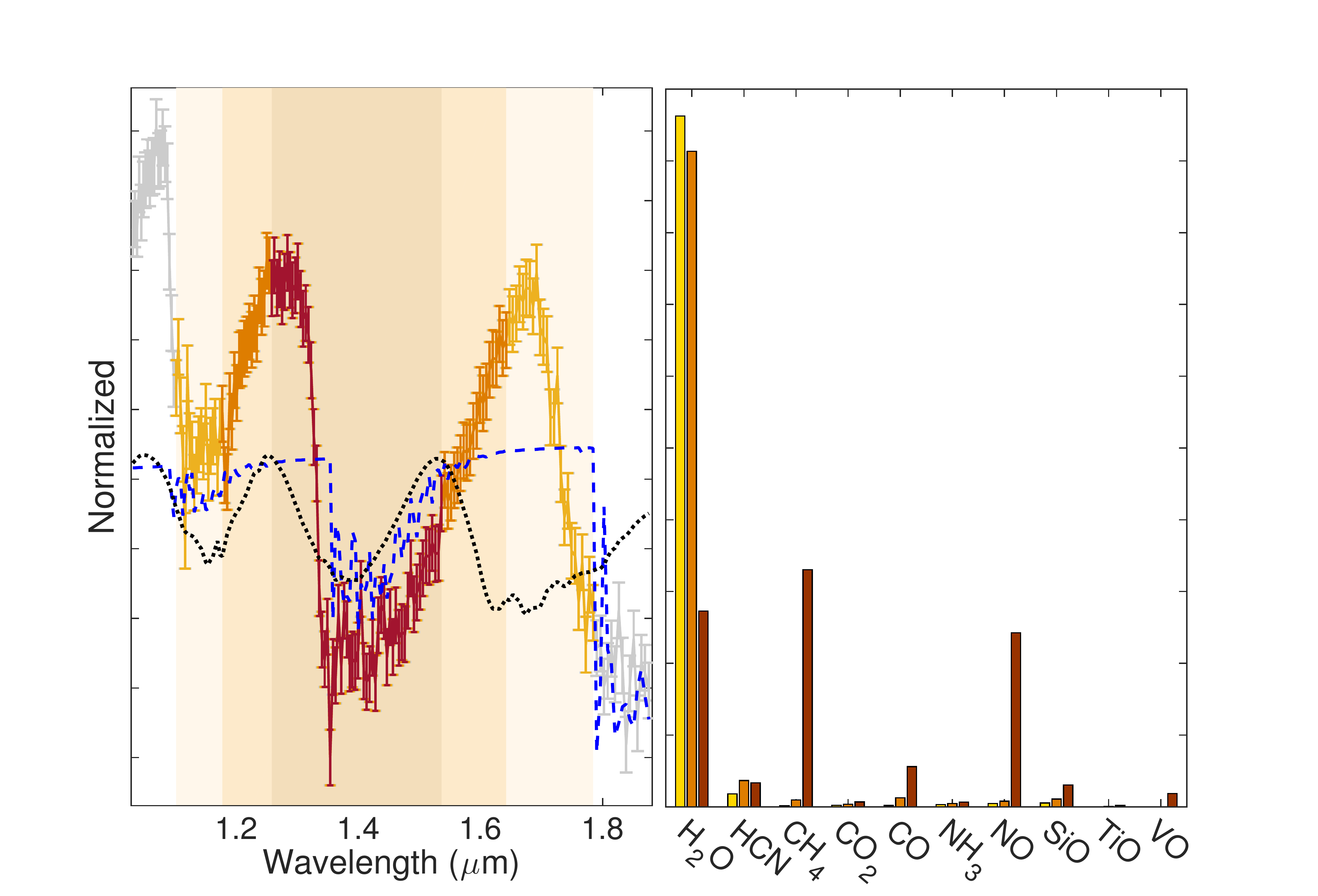}
\caption{Similar to figure~\ref{fig:identification}; Left shows input spectrum at S/N = 20 for a normalised water spectrum in the HST/WFC3 G141 grism passband (yellow, 1.1 - 1.8 $\mu$m). Darker colour shading represents progressively smaller passbands for which the recognition was performed. Blue dashed and black dotted lines show normalised spectra of NO and CH$_{4}$ respectively. Right shows the corresponding detection probability per molecule for the varying wavelength ranges. Water is readily recognised to be the main trace gas component but for the smallest bandpass considered where H$_{2}$O, CH$_{4}$ and NO are assigned roughly equal probabilities. As can be seen in the left plot, H$_{2}$O, CH$_{4}$ and NO normalised spectra all have very similar features when only the most restricted (darkest shaded) spectral range is considered. \label{fig:waverestrict}}
\end{figure}

Whereas it is more adequate to train the DBN with instrument specific resolutions and wavelength ranges, e.g. for HST/WFC3, JWST/MIRI \& JWST/NIRSPEC, it is an intriguing exercise in itself to explore the effect of incomplete wavelength ranges on \robert's ability to recognise molecular species. As stated previously, in this example \robert~was trained on a wavelength grid ranging from 1-20$\mu$m with a constant resolution of 300. Figure~\ref{fig:waverestrict} shows the normalised water-only emission spectrum for the HST/WFC3 G141 grism wavelength range (yellow spectrum). The remaining spectrum outside the wavelength range considered is padded with zeros on both sides. \robert~is clearly able to identify water as dominant trace gas. We now consider increasingly restrictive wavelength ranges until the clear water detection breaks down at the 1.26 - 1.53$\mu$m bandpass and \robert~attributes nearly equal probabilities to H$_{2}$O, CH$_{4}$ and NO. Whilst initially surprising, upon closer inspection all three molecular species have strong overlapping features in this wavelength range (blue and black lines in figure~\ref{fig:waverestrict} show the normalised spectrum of NO at $1\times10^{-2}$ and CH$_{4}$ at $1\times10^{-4}$ respectively) and a `visual' separation of molecules becomes very difficult. 

We now investigate the effect of resolution miss-matches between the observed data and the resolution with which the DBN is trained. As expected, downsampling from a higher resolution to the DBN resolution does not impair recognition efficiency. The effect of upsampling, i.e. interpolating the observed spectrum to the resolution of the DBN, is more case dependent. We find no degradation of the recognition efficiency upsampling broad absorbing species such as H$_{2}$O or CH$_{4}$ from resolutions as low as R = 30 to the native resolution of the DBN. Here, the interpolation simply adds noise to the spectrum against which the DBN is very robust. Generally speaking, all molecules can be identified unless their features are strongly undersampled. Trace gases with more narrow emission/absorption bands (e.g. CO, NO) are hence more strongly affected.  For the molecular mixtures considered here, we find a conservative lower limit of R $\sim$ 25 (constant with $\lambda$) below which feature detection becomes difficult. It should be noted that a strongly undersampled spectrum will always be difficult to interpret independently of the methodology used.

\section{Dreaming of atmospheres}
\label{sec:dreaming}

\begin{figure}
\center
\includegraphics[width=1.05\columnwidth]{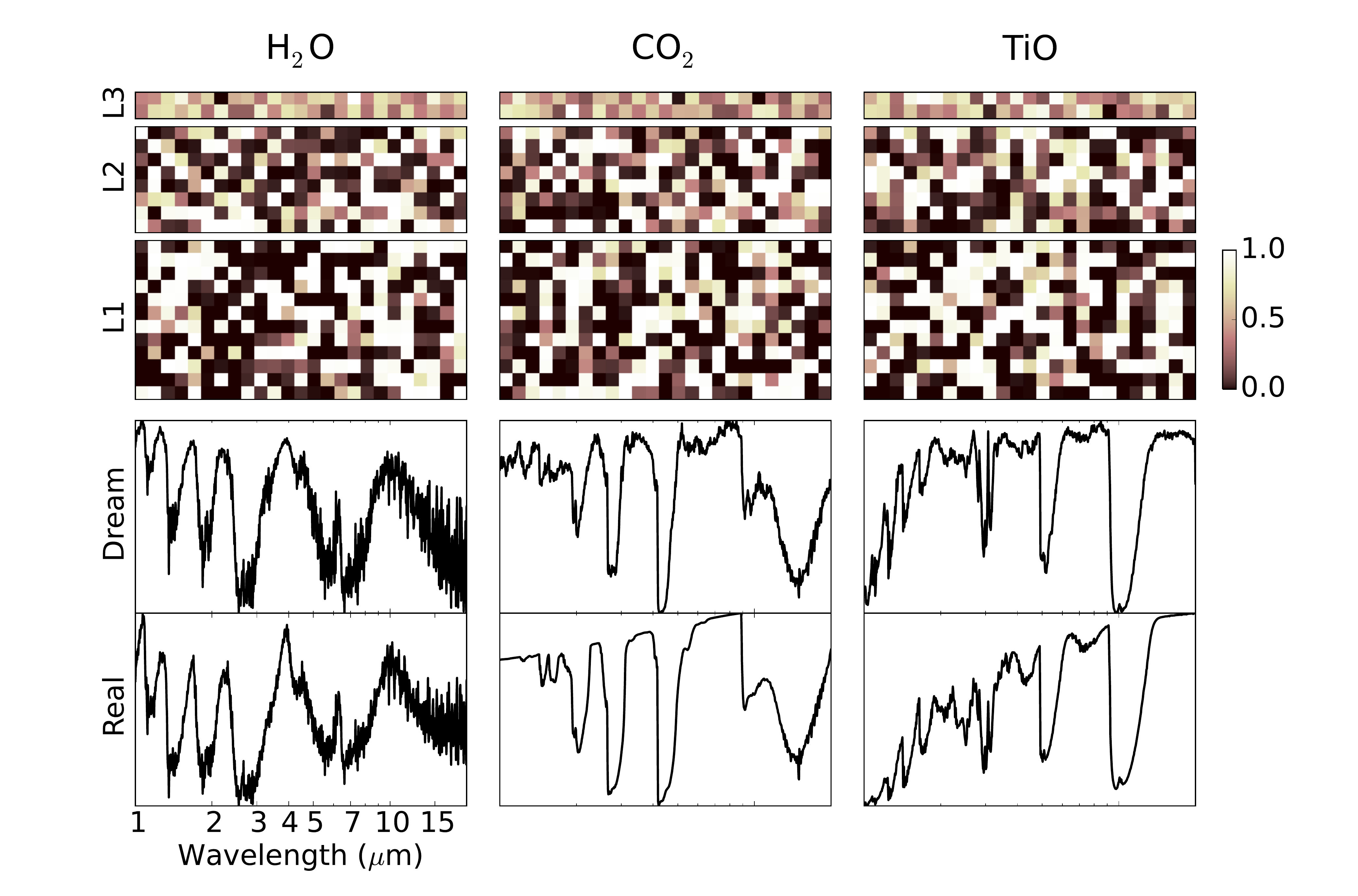}
\caption{Spectral reconstruction (or `dreaming') of three molecules H$_{2}$O, CO$_{2}$ \& TiO. Top three panels show neuron activations for the bottom (L1) to top (L3) Restricted Boltzmann Machine layers. Bottom two rows show normalised  H$_{2}$O, CO$_{2}$ \& TiO spectra reconstructed by the neural network and real data examples as comparison. The similarities between `dreamed' and real spectral features are striking. This indicates a good representation of molecular features in the neural network. \label{fig:dreaming}}
\end{figure}

When \robert~is used for recognition purposes we set the visible units to the values of the input spectrum and propagate the network forward (i.e. upwards) to obtain a classification label. Another approach to qualitatively check the convergence quality of the DBN is to reverse the network and propagate the network weights backwards (i.e. downwards) starting from a label. In other words, we activate the, say, H$_{2}$O label and \robert~will return what it `thinks' are the defining features of a water spectrum. This backwards propagation is commonly referred to as `dreaming' in the machine learning literature. Figure~\ref{fig:dreaming} shows dreams of three molecules, H$_{2}$O, CO$_{2}$ and TiO. We compare these dreams with real, normalised spectra with abundances of $1\times10^{4}$ underneath. The likeness of the dreamed spectra with real data is striking. L1, L2 and L3 represent the neural activations of the bottom, middle and top RBMs respectively. We find the neural activations in the dream state to be a useful indicator of the sparsity (i.e. number of units set to or close to zero) of the neural network and find networks with $\sim$10$\%$ average sparsity to yield the most accurate spectral reconstructions. 

\section{Discussion}
\label{sec:discussion}

The size of the DBN is an important factor to be considered, \robert~consists of three RBMs \'a 500, 200 and 50 neurons from bottom to top respectively. We find a three layer DBN to work best but also find that networks with too many neurons per layer, particularly in the upper levels, lead to noisy reconstructions, low maximum likelihoods, and a poorer recognition performance. We attribute this effect to a high level of redundancy in the network, which introduces noise. As described above, by inspecting the neural activations during the dream state of \robert, we can measure the sparsity of individual layers for individual states (i.e. molecule activations). Tests have shown that $\sim$10$\%$ in sparsity averaged across activation states produces the most robust and highest S/N networks. Smaller, simpler networks run the risk of not being able to differentiate between molecules correctly. 

As stated previously, \robert~has only been trained on spectra containing one trace gas at a time. Despite this obvious limitation, we show in section~\ref{sec:identification} that \robert~is indeed able to identify  mixtures of molecules, though caveats to this capability should be mentioned. Similar to inspecting a spectrum by eye, \robert~is able to identify mixtures if the trace gas signatures are very different to one another (e.g. H$_{2}$O and CO, figure~\ref{fig:dreaming}) or if sufficient wavelength coverage is provided (e.g. CH$_{4}$ and H$_{2}$O, figure~\ref{fig:identification}). The DBN struggles whenever either too little wavelength coverage is available (e.g. figure~\ref{fig:waverestrict}) or the secondary trace gas is of an order of magnitude less abundant than the primary absorber/emitter, i.e. secondary signatures imprint themselves as noise on the main absorber/emitter.  

Though some of these limitations are fundamental (i.e. insufficient wavelength coverage, too low S/N, etc.), future work will investigate the use of convolutional deep belief networks \citep[e.g.][]{Lee:2011kq} to boost recognition accuracy by learning the localised correlations in the observed spectra. Additionally, an updated supervised learning cost-function is imaginable where not the identification of a single trace gas is rewarded but instead a `best ranking' of groups of molecules. 

As pre-selector to the \taurex~retrieval suite, \robert~will provide rankings of the most likely molecules to be considered in the quantitative retrieval. This is an iterative process with the retrieval models increasing in complexity, from the simplest atmospheres (containing only the few most likely molecular absorbers/emitters detected by \robert) to more complex models (containing less likely opacities). The Bayes factor is the measure of convergence here \citep{waldmann15}. In future implementations of \robert, online-learning will become important after its initial training phase is complete. With each new data set, \robert~will be able to update and improve its DBN, taking the \taurex~results as labeled training set. Such an application is particularly suited as part of a larger data reduction/analysis pipeline for future large scale ground and space-based surveys. 

\section{Conclusion}

In this paper we present the use of Deep belief networks in the identification and classification of exoplanetary emission spectra. 
We have shown that DBNs are well suited to identifying molecular signatures in extrasolar planet spectra. They are very robust to low S/N ratios and are able to identify trace gases even when  wavelength ranges are strongly restricted compared to the initial training setup. This property is important as training a DBN is relatively computationally intensive and hence one would ideally want the trained DBN to be as universally applicable as possible. Their ability to abstract and generalise non-linear systems very effectively, makes DBNs an ideal tool for qualitative `pre-selection' of parameter spaces for spectral retrieval applications.



 \section*{Acknowledgements}
IPW thanks G. Tinetti, R. Varley, M. Rocchetto, A. Tsiaras \& G. Morello for useful discussions. This work was supported by the ERC project 617119 (ExoLights).

\bibliographystyle{apj}
\bibliography{taurex-lib,taunet-lib}

\end{document}